\documentclass[twocolumn,showpacs,preprintnumber,amsmath,amssymb,floatfix,prb]{revtex4}

\newcommand{\ket}[1]{\left | #1 \right \rangle}                      

\newcommand{\mean}[1]{\left \langle #1 \right \rangle}

\newcommand{\un}[1]{\underline{#1}}

\newcommand{\off}{\! \textit{off}}
\newcommand{\phem}[1]{#1}

\usepackage{graphicx}

\begin{document}

\preprint{APS/123-QED}

\title{Designable electron transport features in one-dimensional arrays of metallic nanoparticles:
Monte Carlo study of the relation between shape and transport}

\author{Stefan Semrau}
 \author{Herbert Schoeller}%
\affiliation{%
Institut f\"ur Theoretische Physik A, RWTH Aachen
}%

\author{Wolfgang Wenzel}
\affiliation{%
Institut f\"ur Nanotechnologie, FZ Karlsruhe
}%

\date{\today}

\begin{abstract}
We study the current and shot noise in a linear array of metallic nanoparticles taking explicitly into consideration their discrete electronic spectra. Phonon assisted tunneling and dissipative effects on single nanoparticles are incorporated as well. The capacitance matrix which determines the classical Coulomb interaction within the capacitance model is calculated numerically from a realistic geometry. A Monte Carlo algorithm which self-adapts to the size of the system allows us to simulate the single-electron transport properties within a semiclassical framework. We present several effects that are related to the geometry and the one-electron level spacing like e.g.~a negative differential conductance (NDC) effect. Consequently these effects are designable by the choice of the size and arrangement of the nanoparticles.
\end{abstract}

\maketitle

Programmable self-assembly is one of the most promising bottom-up approaches to the synthesis of nano-electronic devices. For this technique a \phem{template} is needed which determines the desired device shape.  It has turned out that \phem{DNA} is ideal for this purpose\cite{Niemeyer04}: because of the highly selective binding of single-stranded DNA to another strand with complementary bases, it can be conformed to a variety of geometrical shapes\cite{Zhang94}. By decorating the DNA with metal nanoparticles conducting material can be created. The attachment of gold nanoparticles with a selective size to surfaces \cite{Noyong03}, certain biopolymers \cite{Berven02} and to DNA \cite{Noyong03b} has already been achieved. 
\begin{figure}[h]
\centering
\includegraphics[width=0.45\textwidth]{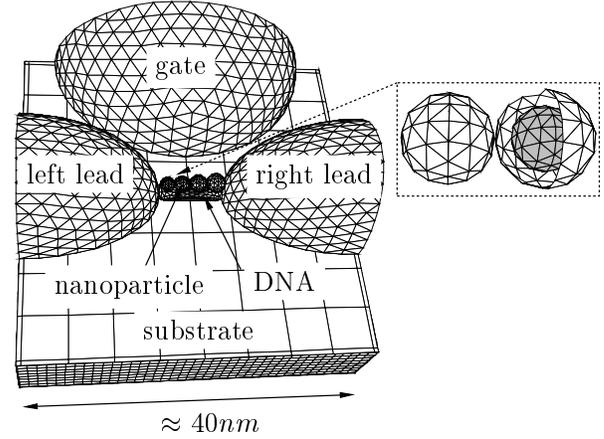}
\caption{\label{fig:geomex} Wireframe of a typical considered geometry. Adopting the capacitance model we assume that gate, leads and nanoparticles are ideal conductors while the other parts of the system are modelled as dielectrics. Gate and leads are quarter ellipsoids. The substrate is a dielectric cuboid with a relative permittivity of $3.9$ (silicon dioxide). For the nanoparticles we assume a spherical geometry. The ligand shell which is necessary to bind the nanoparticles to the DNA is modelled as a concentric dielectric shell of $0.3nm$ thickness with a relative permittivity of $3$. This is shown in the inset of Fig.~\ref{fig:geomex}: the metallic nanoparticle itself is depicted in grey while the space between the particle and the outer sphere is filled with the dielectric. Finally the DNA is modelled as a dielectric cylinder with a diameter of $2 nm$ and a relative permittivity of $3$. (The figure was made with ``FastCap''\cite{White91}.)}
\centering
\end{figure}
In this paper we focus on linear arrays of nanoparticles -- also called (quantum) dots. This geometry is interesting for applications e.g. since its electron transport properties are robust against unintentional fluctuationg background charges \cite{Wasshuber01}. On the other hand, fundamental transport phenomena can be observed in such systems\cite{Bylander05, Averin05}. In our case, the nanoparticles are attached to a DNA strand via a functionalizing ligand shell, see Fig.~\ref{fig:geomex} for a wireframe. The DNA strand, which resides on a dielectric substrate, is stretched between two macroscopic metal leads, serving as electron reservoirs, to which different potentials can be applied. A third metal lead serves as a gate.

We determine the current originating from the tunneling of single electrons and the corresponding shot noise as functions of various parameters like e.g. the applied gate and bias voltage, the temperature and the strength of dissipative effects. Especially the geometry of the array plays a major role since it is intended to design the electron transport properties by controlling the shape of the device. From existing studies of arrays of metallic nanoparticles with small capacitances and high junction resistances \cite{Amman89,Amman89b,Geigenmueller89,Bakhvalov89,Middleton93,Melsen97,Nguyen01} it is known that such systems exhibit \phem{nonlinear current-voltage characteristics} -- also called IV characteristics -- which can be used for interesting electronic applications\cite{Likharev99}. Typically the IV characteristics for an array with $Z$ nanoparticles have the form:
\begin{equation}
I(V) \propto \left ( V/V_{T} - 1 \right )^{a} \Theta \left (V - V_{T} \right ) \quad , \quad V_{T} \propto Z^{b} E_{c}
\end{equation}
Here $V_{T}$ is the \phem{threshold voltage}, which borders the region of the \phem{Coulomb blockade} -- in a certain bias voltage range no current flows due to the Coulomb interaction between charged nanoparticles, $E_{c}$ is the \phem{charging energy}, which is a measure for the energy necessary to charge a nanoparticle with an additional elementary charge.

In the studies mentioned above, the electronic transport is addressed within the so-called \phem{orthodox theory}. Our treatment differs substantially from that theory in two aspects: firstly it is not needed to assume that the mean occupation of the electronic levels on the nanoparticles is given by a Fermi distribution. Instead it is in general determined both by the tunneling kinetics and \phem{dissipative effects} on the nanoparticles. If the latter dominate, they drive the mean configuration to a Fermi distribution which is the limit used in the orthodox theory. But within our model the strength of the dissipation can also be weak. We focussed on the latter case in which the mean occupation on a nanoparticle is determined by the tunneling kinetics alone. Secondly, in the orthodox theory the electronic spectrum is assumed to be continuous. In this paper, however, the nanoparticles are considered to be so small that the level spacing can have the same order of magnitude as the thermal energy $k_{B}T$ or the charging energy $E_{c}$. So we take into account explicitly the \phem{discrete nature of the electronic spectrum} of the nanoparticles. Doing so, the number of many-particle states that may take part in transport is so huge that the problem completely defies an analytical solution. Instead we use a \phem{Monte Carlo method}  to determine the electron transport properties. The algorithm used here is different from the one used in previous studies\cite{Amman89,Amman89b,Nguyen01}: it copes with the hugeness of the state space and is partially self-adapting to the size of the examined system. As in the orthodox theory, tunneling is treated as a perturbation while the Coulomb interaction between charged nanoparticles is taken into account nonperturbatively within a \phem{capacitance model}. In this model the capacitance matrix determines the classical effects of Coulomb interaction and therefore the transport properties of the system. To elucidate the relation between the shape of the device and its transport properties the capacitance matrix is extracted numerically from realistic array geometries.

With our model we are able to identify one-electron levels and study the interplay of the charging energy and the one-electron level spacing. We will further demonstrate the impact of dissipation on the IV characteristics.  We will present several effects which are related to the geometry and are therefore designable: the IV characteristics are strongly asymmetric due to asymmetric capacitance matrices and the level spacing varying over the array. We will show that in the investigated geometry the conductance of the ohmic parts of the IV curves may unexpectedly  rise with a growing array length. A designable NDC effect occurs if finite electronic spectra on the nanoparticles are considered.

\section{\label{sec:model}Model}
\subsection{\label{subsec:hamil}Model system}
\begin{figure*} 
\centering
\includegraphics[width=\textwidth]{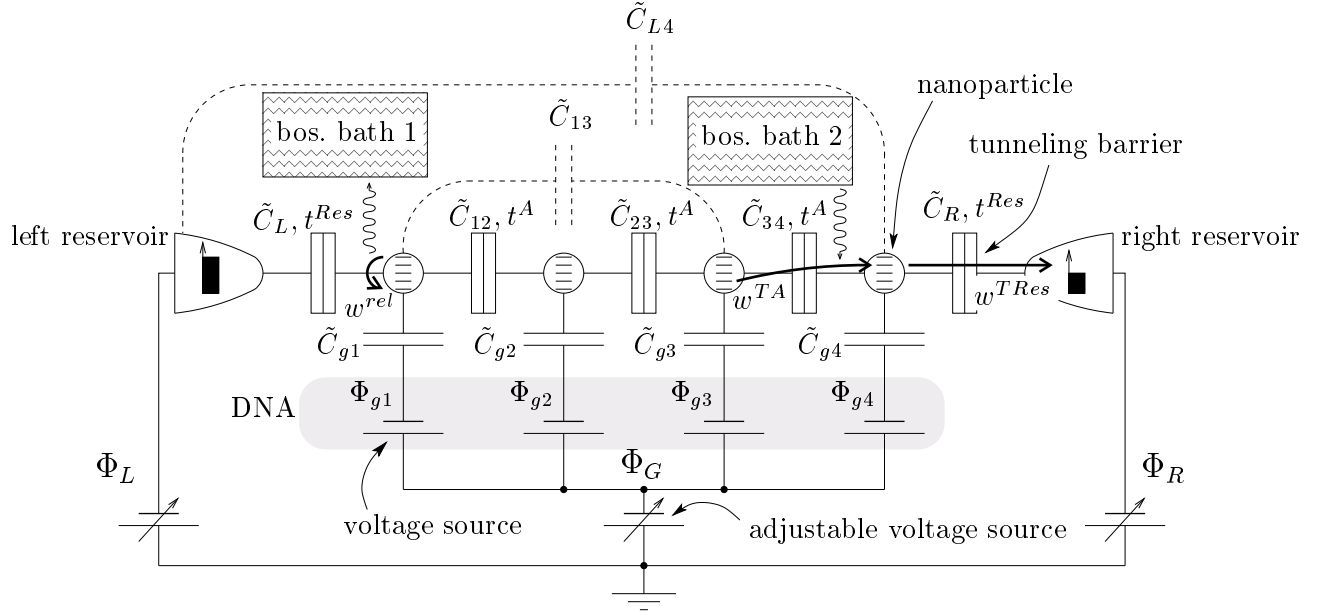}
\centering
\caption{\label{fig:circuit} Model system}
\end{figure*}

In Fig.~\ref{fig:circuit} a pictorial representation of the model system is shown. The system is modelled by a \phem{tight-binding tunneling Hamiltonian for spinless electrons}. The free part consists of the electronic spectra of the reservoirs and the nanoparticles as well as the classical Coulomb interaction. The reservoirs shall consist of non-interacting electrons with a continuous, homogenous, infinite spectrum and they shall be in thermal equilibrium at all times. Their occupation numbers accordingly obey a Fermi distribution. The one-electron spectra on the $Z$ nanoparticles are considered to be discrete and on dot $i$ we consider explicitly $Z_{i}$ levels with level spacing $\Delta \varepsilon_{i}$. The Coulomb interaction is incorporated by the capacitance model which is detailed in the next section. We consider mutual capacitances between nearest neighbours like e.g. $\tilde{C}_{12}$ as well as between distant conductors like e.g. $\tilde{C}_{L4}$. Potentials can be applied to the reservoirs ($\Phi_{L},\Phi_{R}$) and a gate ($\Phi_{G}$). The uncontrollable influence of the DNA creates background charges on the dots (indicated by the random potentials $\Phi_{gi}$). 

The perturbation comprises three types of transitions:
\emph{Transitions within the array} We consider phonon assisted tunneling between the one-electron levels of nearest neighboring dots ($w^{TA}$). The tunneling matrix element $t^A$ is assumed to be equal for all tunneling transitions within the array. The phonon assisted tunneling stems from a linear coupling of a bosonic bath (bosonic bath 2) and the electronic degrees of freedom in the array\cite{Nazarov92,Devoret94}. Here the bosonic bath shall model phonons in the substrate.

\emph{Transitions between array and reservoir} We include tunneling between the continuous reservoirs and the one-electron levels of the outermost nanoparticles where no phonon assistance is considered here ($w^{TRes}$). The tunneling matrix element $t^{Res}$ for these transitions is considered to be equal for both reservoirs but generally different from $t^{A}$.

\emph{Transitions on a single dot} These transitions among the one-electron levels on a single dot ($w^{rel}$) can be justified microscopically by a Fr\"ohlich-Hamiltonian\cite{Mahan90} (coupling to bosonic bath 1). 

We assume that the tunneling matrix elements have phases which fluctuate randomly due to slight temporary changes of the array geometry. Implicit summation over these phases prohibits any first order contributions of the tunneling matrix elements to the perturbation expansion. 

\subsection{\label{subsec:transenerg}Transition energies}
The transition rates which are given in the next section are determined by the  transition energies,  i.e. the changes in energy of the array that accompany the transitions of an electron from one level to another. It is composed of a change in electrostatic and one-electron energy. The former is computed within the capacitance model: the reservoirs, the gate and the dots are treated as macroscopic conductors with potentials $\Phi_{i}$ and total charges $Q_{i}$. We call the gate and the reservoirs \phem{voltage nodes} (with potentials $\un{\Phi}_{V}$ and charges $\un{Q}_{V}$) since their potential is fixed, the dots are called \phem{charge nodes} (with potentials $\un{\Phi}_{C}$ and charges $\un{Q}_{C}$) since their charge is known and their potentials have to be determined\cite{Wasshuber01}. The potentials and the charges are related by the capacitance matrix $C$ 
\begin{equation}
\left(
\begin{array}{c}
\un{Q}_{C} \\
\un{Q}_{V} \\
\end{array}
\right)
= C
\left(
\begin{array}{c}
\un{\Phi}_{C} \\
\un{\Phi}_{V} \\
\end{array}
\right)
=
\left(
\begin{array}{cc}
C_{c} & C_{b}^{\top} \\
C_{b} & C_{a}
\end{array}
\right)
\left(
\begin{array}{c}
\un{\Phi}_{C} \\
\un{\Phi}_{V} \\
\end{array}
\right)
\label{eq:cap3}
\end{equation}
The relation between the mutual capacitances $\tilde{C}_{ij}$ indicated in Fig.~\ref{fig:circuit} and the elements of the capacitance matrix $C_{ij}$ is given by $C_{ij} = \delta_{ij} \sum_{k=0}^{Z} \tilde{C}_{jk} - \tilde{C}_{ij}$. The electrostatic energy can be written as:
\begin{equation}
E_{el} = \frac{1}{2} \left ( \un{Q}_{C} + \un{Q}'_{C} + \un{Q}^{bg}_{C}\right )^{\top} C_{c}^{-1} \left ( \un{ Q}_{C} + \un{Q}'_{C} + \un{Q}^{bg}_{C} \right )
\label{eq:elenergy}
\end{equation}
where $\un{Q}'_{C} : = - C_{b}^{\top} \un{ \Phi}_{V}$ is the polarization charge induced by the potentials $\un{\Phi}_{V}$ on the voltage nodes and $\un{Q}^{bg}_{C}$ are fixed background charges which account for the electrostatic influence of the DNA and unknown fabricational details.  The  potentials on the charge nodes are then:
\begin{equation}
\un{\Phi}_{C} =  C^{-1}_{c} \left (\un{Q}_{C} + \un{Q}'_{C}  + \un{Q}^{bg}_{C}  \right )
\end{equation}
Furthermore we have to take into account the electronic spectra of the nanoparticles: the one-electron energy $\varepsilon_{li}$ of level $l$ on dot $i$ is $ \varepsilon_{li} = \varepsilon_{Fi} + l \cdot \Delta \varepsilon_{i}$ with the Fermi energy $\varepsilon_{Fi}$ on dot $i$.  We find the following transition energies\footnote{We follow the considerations of Peter Hadley, presented on \\ http://qt.tn.tudelft.nl/$\sim$hadley/set/electrostatics.html }:

\emph{Transitions within the array} from level $l$ on dot $i$ to level $l'$ on dot $i \pm 1$
\begin{align}
\Delta E^{\pm,i,l'l} &= l' \cdot \Delta \varepsilon_{ i \pm 1} - l \cdot \Delta \varepsilon_{i}   \nonumber \\
-e &\left(\Phi_{Ci \pm 1} - \Phi_{Ci} \right )  +  \frac{e^2}{2} \left ( C^{-1}_{ii}  - 2 C^{-1}_{i i \pm 1}  + C^{-1}_{i \pm 1 i \pm 1} \right )
\label{eq:deltaEarr}
\end{align}
Note that the dependence of the potentials $\un{\Phi}_{C}$ on the charges $\un{Q}_{C}$ is implied and  that we include the constant terms $\varepsilon_{Fi}$ in the definition of $\un{\Phi}_{C}$, i.e. we treat them like background charges:
\begin{equation}
\un{\Phi}_{C} = C^{-1}_{c} \left (\un{Q}_{C} + \un{Q}'_{C}  + \un{\tilde{Q}}^{bg}_{C}  \right ) \text{ with }  \un{\tilde{Q}}^{bg}_{C}  =  \un{Q}^{bg}_{C}  - \frac{1}{e} C_{c} \un{\varepsilon}_{F} \label{eq:bgcharge}
\end{equation}
where  the vector $\un{\varepsilon}_{F} $ contains the constants $\varepsilon_{Fi}$.

\emph{Transitions between array and reservoir} from/to lead $\alpha$ into/out of level $l$ on the neighbouring dot
\begin{equation}
\Delta E^{\pm, \alpha,l} = \pm  l \cdot \Delta \varepsilon_{\beta} \pm (-e) \left( \Phi_{C\beta} -\Phi_{V \alpha} \right) + C^{-1}_{\beta \beta} \frac{e^2}{2} \label{eq:deltaElead}
\end{equation}
where $\beta=1$ for $\alpha=L$ (left lead) and $\beta=Z$ for $\alpha=R$ (right lead). (Note that the nanoparticles are numbered consecutively from left to right.)

\emph{Transitions on a single dot} from level $l$ to level $l'$ on dot $i$ only change the one-electron energy:
\begin{equation}
\Delta E^{i,l'l} = \left (l' - l \right ) \cdot \Delta \varepsilon_{i} \label{eq:deltaErel}
\end{equation}
It is convenient to express the transition energies in this way since we have to calculate the potentials $\un{\Phi}_{C}$ only once at the beginning of the simulation with \eqref{eq:bgcharge} and then update them after each transition which is computationally cheap. 

Obviously, the behaviour of the system depends strongly both on the values of the mutual capacitances\cite{Whan96} and the one-electron level spacings. Concerning the former we determine the capacitance matrix numerically with the help of ``FastCap''\cite{White91} using the geometry shown in Fig.~\ref{fig:geomex}. Only qualitatively we estimate the magnitude of the one-electron level spacing of a single, neutral, isolated nanoparticle\cite{Halperin86}: 
\begin{equation}
\label{eq:levelspacing}
\Delta \varepsilon_{i} =  \frac{3 \hbar^{2} \pi}{2m_{e} \left ( 3 \pi^{2} n_{i} \right )^{\frac{1}{3}}} \cdot \frac{1}{r_{i}^{3}} =  \frac{c}{r_{i}^{3}} \;,\; c \approx 0,30 \cdot 10^{-28} eVm^{3}
\end{equation}
where $n_{i}$ is the electron density, $r_{i}$ the nanoparticle radius and the numerical value for $c$ was calculated for bulk gold. DFT calculations \cite{Barnett99} and experiments \cite{Zhang04} suggest that this estimation can at least reproduce the correct order of magnitude.

\section{\label{sec:method}Method}
\subsection{\label{subsec:transport}Transport theory}
We assume that the tunneling rates defined below are much smaller than $(k_{B}T)/ \hbar$ so that we can treat the tunneling part of the Hamiltonian as a perturbation (\phem{weak coupling regime}) and consider only the lowest non-vanishing order of the perturbation expansion (\phem{sequential tunneling regime}). The reservoir and bath degrees of freedom of the density matrix are traced out which results in the \phem{reduced} density matrix. We neglect its non-diagonal elements, which is justified if the broadening of the levels due to tunneling is small compared to the level spacing. The diagonal elements $P_{s}$ can then be interpreted as the probabilities of the array states $\ket{s}$ where the array state $s$ is given by all occupation numbers $n_{li}$ of level $l$ on dot $i$: $ \ket{s} = \ket{ \left\lbrace  n_{li} \right\rbrace_{li}  } $. The rates of electron transfer are calculated in golden rule approximation, thereby assuming that transport is incoherent and no coherent eigenstates are formed which stretch over the whole array, comparable to molecular orbitals. This is justified since in reality there are certainly processes which destroy the phase coherence -- e.g. slight temporary changes in the geometry of the array.

Under the given assumptions the probabilities $P_{s}$ obey a master equation in the stationary limit\cite{Schoeller97}
\begin{equation}
\dot{P}_{s} = \sum_{s'} \left ( w_{s s'} P_{s'} - w_{s' s} P_{s} \right ) = 0
\label{eq:master}
\end{equation}
with golden rule rates $w_{s's}$ from array state $s$ to array state $s'$. Each transition rate belongs to exactly one of the following sets:

\emph{Transitions within the array} from level $l$ on dot $i$ to level $l'$ on dot $i \pm 1$
\begin{align}
w^{TA} \equiv w^{\pm,i,l',l} &= \Gamma^{A}  P \! \left (\Delta E^{\pm, i, l',l} \right ) \; , \; \Gamma^{A} = \frac{2\pi}{\hbar} \left | t^{A} \right |^{2}\label{eq:warr} \nonumber \\
\text{ with } P(E) &= \frac{1}{1+ e^{\beta E}} \cdot \frac{2}{ \pi} \cdot \frac{\Gamma}{ E^{2} + \Gamma^{2}} \quad , \quad \beta = \left ( k_{B}T \right )^{-1}
\end{align}
where $t^{A}$ is the tunnneling matrix element within the array. The function $P(E)$ is the probability per energy for the exchange of energy $E$ with the bosonic bath and therefore has to be normalized and must fulfill the condition of detailed balance\cite{Nazarov92} $ \int_{-\infty}^{\infty} dE P(E) = 1 \quad, \quad P(-E) = e^{-\beta E} P(E)$. The latter property stems from the nature of the bosonic bath: while it is always possible to emit energy, there must be excited bosons in the bath if energy shall be absorbed. The $P(E)$ function given above is an assumption which is preferably simple and has the required properties. 

\emph{Transitions between array  and reservoir} from/to lead $\alpha$ into/out of level $l$ on the neighbouring dot
\begin{equation}
w^{TRes} \equiv w^{\pm,\alpha,l} = \Gamma^{\alpha} f^{\pm} \! \left (\pm \Delta E^{\pm,\alpha,l} \right ) \; , \; \Gamma^{\alpha} =  \frac{2 \pi}{\hbar} \left | t^{Res} \right |^{2} d_{\alpha}\label{eq:wlead}
\end{equation}
where $t^{Res}$ is the tunnneling matrix element between a reservoir and the neighbouring dot, $d_{\alpha}$ is the density of states in the reservoir $\alpha$ and $f^{+}(E) = (e^{\beta E} +1)^{-1} $, $f^{-} \equiv  1 - f^{+}$.

\emph{Transitions on a single dot} from level $l$ to level $l'$ on dot $i$
\begin{align}
w^{rel} \equiv w^{i,l'l} = \Gamma^{rel} & ( \Theta ( \Delta E^{i,l'l}  ) g (\Delta E^{i,l'l} )   \nonumber \\
    &+  \Theta ( -\Delta E^{i,l'l} )  ( 1+ g ( - \Delta E^{i,l'l}  )  )  )
\end{align}
where $\Gamma^{rel}$ is the inverse relaxation time,  $g(E) = (e^{\beta E} -1)^{-1}$ and $ \Theta(E)$ is the Heaviside step function. Since, as in the case of tunneling within the array, these transitions are possible due to the coupling to a bosonic bath, the rates fulfill detailed balance: $
w^{i,l'l}_{\phantom c} = e^{- \beta \Delta E^{i,l'l} } w^{i,ll'}_{\phantom c} \text{ for } \Delta E^{i,l'l} > 0$. If the nanoparticles were isolated, i.e. if there were no transitions among them, it follows from the detailed balance property that in the stationary limit the occupation numbers $n_{li}$ obey a Fermi distribution. So while the tunneling transitions drive the electron distribution out of equilibrium the transitions on single dots effectively cool the electrons.

\subsection{\label{subsec:monte}Monte Carlo algorithm}
\begin{figure}
\centering
\includegraphics[width=0.45\textwidth]{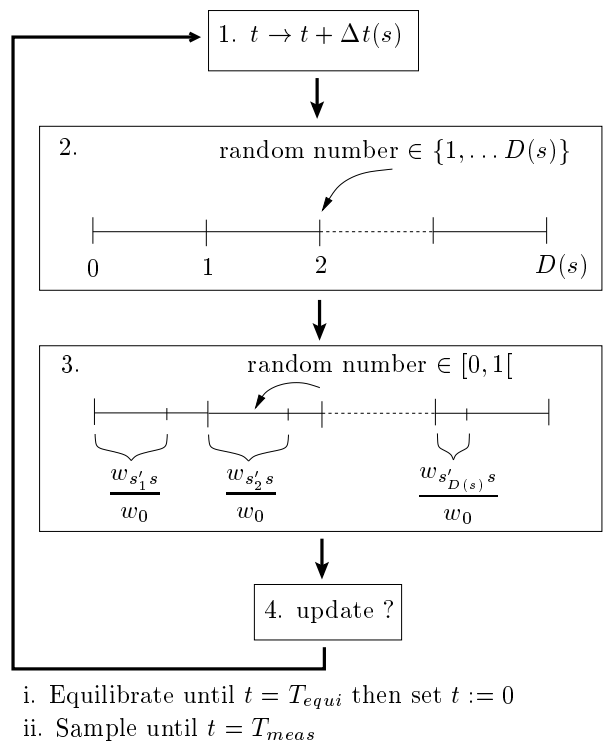}
\centering
\caption{\label{fig:mcalgo2} MC algorithm: 1. Augment time by an amount of $\Delta t(s) = 1 /  \left( w_{0} D(s) \right )$, 2. Choose one of the $D(s)$ transitions with equal probability $1/D(s)$,  in other words choose an \phem{attempt configuration} the system might transit to, 3. Accept the choice with probability $w_{s's}/w_{0}$, 4. If the transition is accepted, update the state of the system, especially the total number of possible transitions $D(s)$. }
\end{figure}

If we consider $Z_{i}$ levels on dot $i$ there are $2^{\sum_{i} Z_{i}}$ possible array states. So the master equation \eqref{eq:master} defies a direct numerical solution except for very small systems. Therefore we employ a Monte Carlo (MC) method to retrieve the quantities of interest: the current and the shot noise. The key idea of the MC method\cite{Honerkamp94} in this context is to discretize time and to get from the transition rate $w_{s's}$ a \phem{transition probability} $\pi_{s's}$ by multiplying the rate with a finite time step $\Delta t(s)$ which may depend on the present system state $s$: $ \pi_{s's}(\Delta t(s)) = w_{s's} \Delta t(s) $ The time step has to be chosen sufficiently small so that for the total probability $\pi_{s}(\Delta t(s))$ to leave the state $s$ it holds
\begin{equation}
\pi_{s}(\Delta t(s)) = \sum_{s'} \pi_{s's}(\Delta t(s))  \leq 1 \quad \forall s \label{eq:transprob}
\end{equation}
The transitions among the array states, which are governed by the probabilities $\pi_{s's}$, constitute a stochastic process which can be simulated with the help of random numbers. In contrast to the MC method used in statistical physics\cite{Binder97} which samples a (grand) canonical ensemble, the system examined here is generally out of equilibrium. In each valid MC algorithm the probability $\pi_{s}(\Delta t(s))$ must be properly represented. We write it as follows:
\begin{equation}
\pi_{s's}(\Delta t(s)) = w_{s's} \Delta t(s) = \frac{w_{s's}}{w_{0}} \cdot \frac{1}{D(s)} \cdot D(s) w_{0} \Delta t(s)
\end{equation}
where $D(s)$ is the number of possible transitions out of a state $s$ and $w_{0}$ is an upper bound for the transition rates which is also called \phem{attempt frequency}. If we choose the time step $\Delta t(s)$ as $\Delta t(s) = 1/ (D(s) w_{0})$, the transition probability reduces to:
\begin{equation}
\pi_{s's}(\Delta t(s)) = \frac{w_{s's}}{w_{0}} \cdot \frac{1}{D(s)} \label{eq:algoprob}
\end{equation}
Note that with the upper choice of the time step the requirement \eqref{eq:transprob} is fulfilled since $w_{s's}  \leq w_{0}$ and the number of nonzero addends is $D(s)$. It is straightforward to imagine the corresponding algorithm if you regard each factor in \eqref{eq:algoprob} as an independent probability, see Fig.~\ref{fig:mcalgo2}.
We found that even though our algorithm needs two random numbers per time step and an attempt transition may be rejected, see step 3 in Fig.~\ref{fig:mcalgo2}, it is convenient since only one rate has to be computed per time step. For systems with many possible transitions it is faster than algorithms which need  the transition rates of all possible transitions in each time step\cite{Amman89,Amman89b}. Furthermore our algorithm has the convenient property to be self-adapting to the system size: since $\Delta t(s) \propto \left ( D(s) \right ) ^{-1} $ the number of performed steps scales with the amount of possible transitions if the runtime of the simulation is fixed.

Despite the optimized algorithm we can take only a finite number of one-electron levels into account, of course.  In most cases, however, we want to model wide bands on the dots, i.e.~we do not want to observe any impact of the finiteness of the electron spectrum. Practically we increase the number of considered levels until the results become steady for the maximum bias voltage which we want to apply. In order to minimize the computing time, i.e. real time, we consider as few levels as possible, of course. Therefore we center the spectra around the highest occupied levels in the ground state (i.e.~the state assumed in equilibrium at zero temperature) since for small bias voltages and low temperatures only low-lying excitations about the Fermi edge appear. To determine the ground state the total energy of the system which includes the electrostatic energy \eqref{eq:elenergy} and the one-electron energies has to be minimized. Due to the discreteness of the charges $\un{Q}_{C}$ this is a non-trivial minimization problem which we do not address here. 

The initial state of a simulation run is always the ground state, i.e.~all spectra are half-filled. Each run starts with an equilibration period $T_{equi}$ in which we let the system evolve, by iterating the algorithm (Fig.~\ref{fig:mcalgo2}), without sampling the assumed states. So the system can reach its stationary state. In the subsequent measurement period $T_{meas}$ the quantities of interest are retrieved. The current can simply be obtained by counting the electrons that are transferred e.g.~between the left lead and the first nanoparticle and dividing by the length of the measurement period $T_{meas}$: $ I_{L} = Q_{L}(T_{meas})/T_{meas} $ where $Q_{L}(T_{meas})$ is the charge that is transferred during the measurement time $T_{meas}$. Note that in the stationary state the current through all tunneling barriers is the same. For a fixed set of parameters the simulation is repeated with different seeds for the random number generator in order to get statistically independent runs. With this ensemble we can determine the statistical standard error of the mean current.  The (zero-frequency) shot noise $S_{I_{L}}(0)$ and the Fano factor $F$ can also be estimated\cite{Korotkov94}:
\begin{align}
S_{I_{L}}(0) &=  \frac{2}{T_{meas}} \left ( \mean{Q_{L}(T_{meas})^{2}}  - \mean{Q_{L}(T_{meas})}^{2} \right ) \nonumber \\
&\quad \quad \quad \quad \quad \quad F = \frac{S_{I_{L}}(0)}{2 e \left \langle I_{L} \right \rangle}
\label{eq:Fano}
\end{align}
where $\left \langle \ldots \right \rangle$ denotes the ensemble average. As for the current, in the stationary limit the shot noise is the same for all tunneling barriers. Note that all given quantities are estimators which become exact in the limit $T_{meas} \rightarrow \infty$.

The validity of our method was checked by comparing our results with the solution of the master equation for a small system. With the same benchmark   sensible values for the simulation parameters (equilibration and measurement time, size of the ensemble) were obtained. Due to the self-adaptive property of the algorithm these parameters are also suitable for bigger systems. For each geometry we increased the number of considered one-electron levels until the IV curves became steady. Furthermore we found that the computing time scales linearly with the number of nanoparticles $Z$ and quadratically with the maximum number of considered levels on a dot ($\max \lbrace \tilde{Z}_{i} \rbrace_{i} $). It is exponentially smaller than the computing time needed for the direct solution of the master equation which scales as $( 2^{\sum_{i}\tilde{Z}_{i}}  )^{3}$. However, note that the MC method is not equivalent to a solution of the master equation: though the current and shot noise may be retrieved from a MC simulation, it is practically impossible to determine all probabilities $P_{s}$ correctly.

\subsection{\label{subsec:charge}Charge states}
To interpret the simulated results we draw a sample out of a single simulation run and look at the probabilities of charge configurations and mean rates among them. To get the probability $P_{C}$ of a certain charge configuration $C \equiv \lbrace Q_{Ci} \rbrace_{i}$, we sum the MC times during which this configuration is assumed and divide by the total MC time. The mean rate $w^{\pm,i}_{C'C}$ through a tunneling junction $i$ from state $C$ to state $C'$ to the right or left respectively is determined by counting the transitions between state $C$ and state $C'$ by tunneling in the given direction through junction $i$ and dividing the sum by the MC time that is spend in the state $C$. The current through a tunneling junction $i$ can then be written as $ I_{i} = \sum_{C,C'} \sum_{\sigma} \sigma w^{\sigma,i}_{C'C} P_{C} \text{ with } \sigma = \pm $. In the stationary state the currents through the junctions are all equal to one another $I_{i} = I_{j} \quad \forall i=j$, so we can write the current as
\begin{equation}
I = \frac{1}{Z+1} \sum_{i} I_{i} =  \frac{1}{Z+1} \sum_{C,C'} P_{C}  w_{C'C} = \sum_{C,C'}  I_{C'C}
\end{equation}
with $w_{C'C} =  \sum_{i,\sigma} \sigma w^{\sigma,i}_{C'C}$ and $I_{C'C} =  \frac{1}{Z+1} P_{C} w_{C'C}$. The \phem{partial current} $I_{C'C}$ from charge state $C$ to charge state $C'$ has a positive sign if it flows from left to right and the opposite sign for the opposite direction. If there are partial currents which flow between the same states -- then they have necessarily opposite directions, we keep only the difference of them so that there is only one net partial current between two charge states.

\section{\label{sec:results}Results}
The following conventions hold for all shown results. The general geometrical setup is the one already shown in Fig.~\ref{fig:geomex}. All energies are normalized to the maximum level spacing  that occurs in the array ($\max \left \lbrace  \Delta \varepsilon_{i} \right \rbrace_{i}$). Instead of voltages (potentials) we use potential energies $eV$ and charges are given in units of $e$. We open the bias voltage window symmetrically  (i.e. $\Phi_{L} =  - \Phi_{R}$) because we do not want to introduce artificially an additional asymmetry. We define $V^{\ast} = (e \Phi_{R})/(\Delta \varepsilon_{max})$ so that the current is positive (i.e. flows from left to right) if $V^{\ast}$ is positive. We divide all calculated golden rule rates by $ \Gamma^{A}  \cdot \max(P(\Delta E))$, so that the maximum rate within the array is equal to $1$. We set the maximum rate between array and reservoir equal to $0.1$ since we assume that the tunnel coupling within the array is stronger. Current and charge are expressed in units of the elementary charge $e$. We normalize the current to the maximum rate in the bulk of the array $ I^{\ast} = (I/e) / (\Gamma^{A} \cdot \max(P(E))) $. The current is defined to be positive if it flows from left to right. The parameter $\Gamma$ of the function $P(E)$ is set to $2$, which is small enough to see individual one-electron levels and big enough to give a sufficiently high current. No error bars appear in the following results since the relative statistical error is always only about 0.1\%, so error bars would not be visible.
\subsection{\label{subsec:general}Generalizations of results for single quantum dots}
\subsubsection{\label{subsubsec:levelspace}Interplay of one-electron level spacing and charging energy}
\begin{figure}
\includegraphics[width=0.49\textwidth]{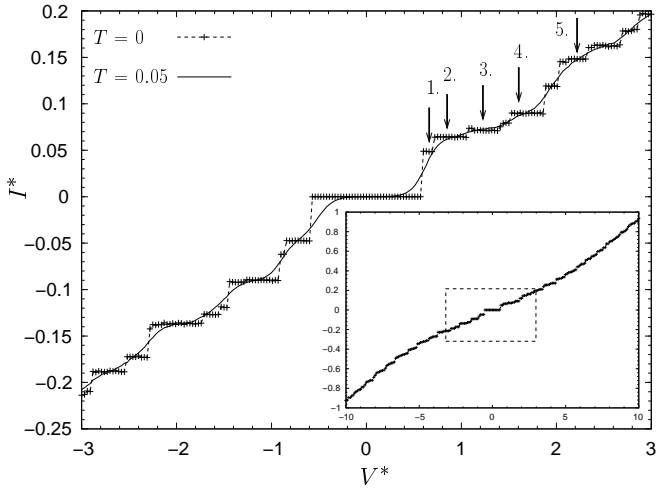}
\caption{\label{fig:arraygeom1a_440_zoom_IV.outs} IV characteristics of an array of 2 nanoparticle with diameters of $1nm$ and $1.2 nm$ respectively. The level spacing of the first equals about $2800 K$ so that a voltage $V^{\ast}$ of 1 equals about $0.24V$. In the inset $V^{\ast}$ ranges from $-10$ to $10$. The region in the dashed rectangle is shown in the big plot: $V^{\ast}$ ranges from $-3$ to $+3$ and the IV curve is shown for $T=0$ and $T=0.05$ . The gate voltage was tuned so that the induced charges on the dots are positive and slightly smaller than an integer value in order to have a small Coulomb blockade region. }
\end{figure}
In a 2 nanoparticle array we investigate the case $\Delta \varepsilon_{i} < |\frac{e^{2}}{2} (C^{-1})_{ii}| $, i.e. the level spacing is smaller than the typical charging energy. In agreement with results for a single dot\cite{Averin91} we find that the level spacing imposes a fine structure on the Coulomb staircase which is related to the charging energy. In Fig.~\ref{fig:arraygeom1a_440_zoom_IV.outs} the corresponding IV characteristics are shown. Looking at the partial currents $I_{C'C}$ defined above we find that for the first 4 steps there is only one relevant transport path in the charge configuration space: $(0,0) \rightarrow (-1,0) \rightarrow (-1,+1) \rightarrow (0,0)$ where the charges are given as differences to the ground state charges. So the first 4 steps must be due to the level spacing. On the 5.~step a second path is relevant: $(0,0) \rightarrow (0,+1) \rightarrow (-1,+1) \rightarrow (0,0)$.   For higher temperatures the fine structure -- i.e.~the features due to the one-electron levels -- is smeared out, see Fig.~\ref{fig:arraygeom1a_440_zoom_IV.outs}, while the typical Coulomb staircase remains: in the middle of each plateau a new transport path becomes available.

\subsubsection{\label{subsubsec:relax}Influence of dissipation}
\begin{figure}
\centering
\includegraphics[width=0.5\textwidth]{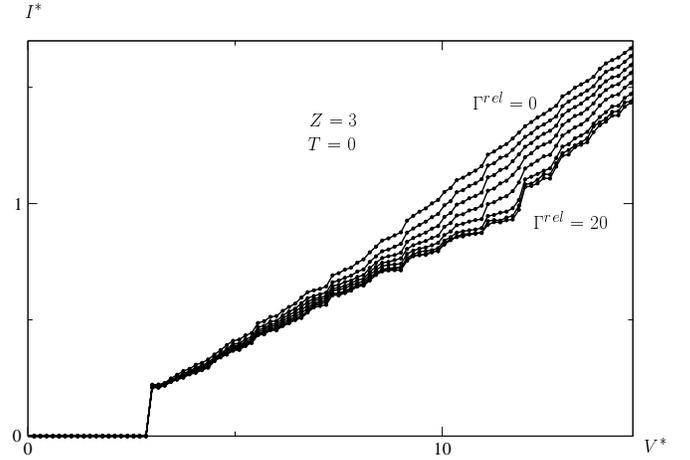}
\caption{\label{fig:arraygeom2a_relaxcomp}Array with 3 nanoparticles, comparison of IV curves for different relaxation rates at a fixed gate voltage. The curves are obtained for values of the relaxation rate prefactor $\Gamma^{rel}$ between $0$ (highest curve) and $20$ (lowest curve) and since the change of the curves' shape is systematic, they are not distinguished.}
\centering
\end{figure}
We consider a 3 nanoparticle array at $T=0$ and fixed gate voltage and study the impact of a finite relaxation rate, see Fig.~\ref{fig:arraygeom2a_relaxcomp}. Generalising results for a single dot\cite{Averin90} we find that without relaxation the electrons overheat and consequently the structures in the IV characteristics are smoothed. Strong relaxation, on the other hand, which effectively cools the electrons, sharpens the steps and decreases the absolute value of the current. The IV curves for intermediate relaxation rates lie between the curves belonging to the extreme cases. These tendencies can also be observed for other gate voltages and other array lengths.  They can be understood by noting that a high relaxation rate keeps the mean occupation of the one-electron levels on a dot close to the equilibrium i.e.~Fermi distribution. For $T=0$ this leads to the formation of a defined Fermi edge on the nanoparticles. That is the reason why the steps in the IV curve become distinct. Electrons above the Fermi energy relax to lower lying levels and the corresponding transition energy is dissipated in the bosonic bath. Such electrons can perform fewer transitions as without dissipation since less energy is available. That is the reason why the current generally decreases with an increasing relaxation rate. 
\subsection{\label{subsec:arrayres}Results uniquely related to the array geometry, designable effects}
\begin{figure}
\centering
\includegraphics[width=0.3\textwidth]{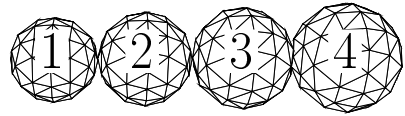}
\caption{Array geometry}
\label{fig:array}
\centering
\end{figure}
We examine arrays of nanoparticles with uniformly growing diameters, see Fig.~\ref{fig:array} for the case of 4 nanoparticles. We choose this special geometry for two reasons: on the one hand the array can be enlarged in a defined way. We start with two nanoparticles and let the number of nanoparticles and therefore the array length grow so that we discover certain features which evolve systematically with increased length. On the other hand it is interesting to combine small and big nanoparticles since they differ strongly both in level spacing and charging energy. In the case of 5 nanoparticles, which is the longest array that is studied here, the diameters of the nanoparticles range from $1nm$ to $1.8nm$ in steps of $0.2nm$. The level spacing of the smallest particle is estimated to be about $0.24 eV$ according to eq.~\eqref{eq:levelspacing} so that a voltage $V^{\ast}$ of $1$ in the following curves equals then $0.24V$. This energy equals a temperature of about $2800 K$ so that we find level spacing related features in the current or shot noise at finite temperatures. Since the size of the nanoparticles increases from left to right, the level spacing decreases in the same direction.

\subsubsection{\label{subsubsec:asym}Strong asymmetry of IV characteristics} 
\begin{figure} 
\centering
\includegraphics[width=0.5\textwidth]{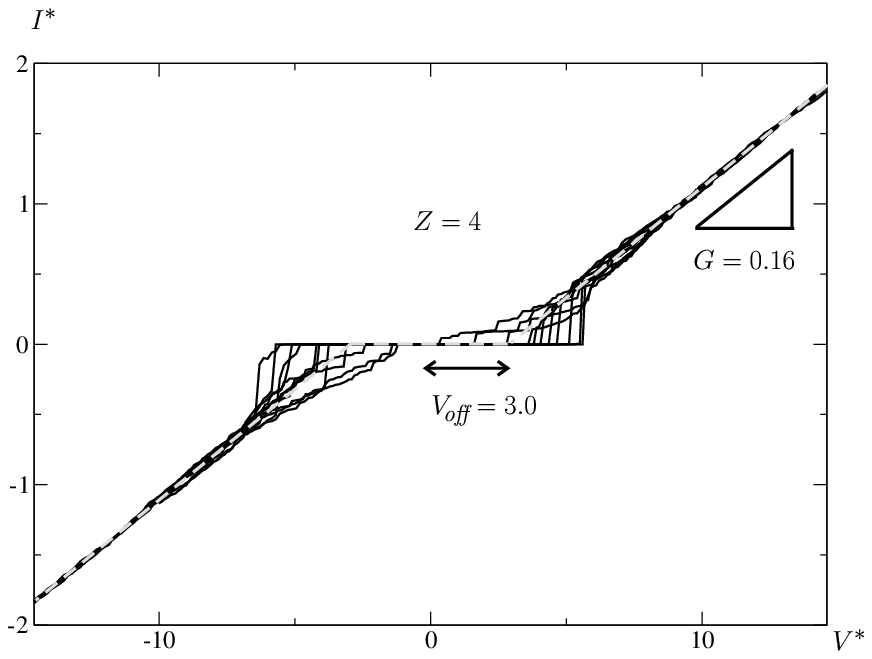}\\[0.2cm]
\includegraphics[width=0.5\textwidth]{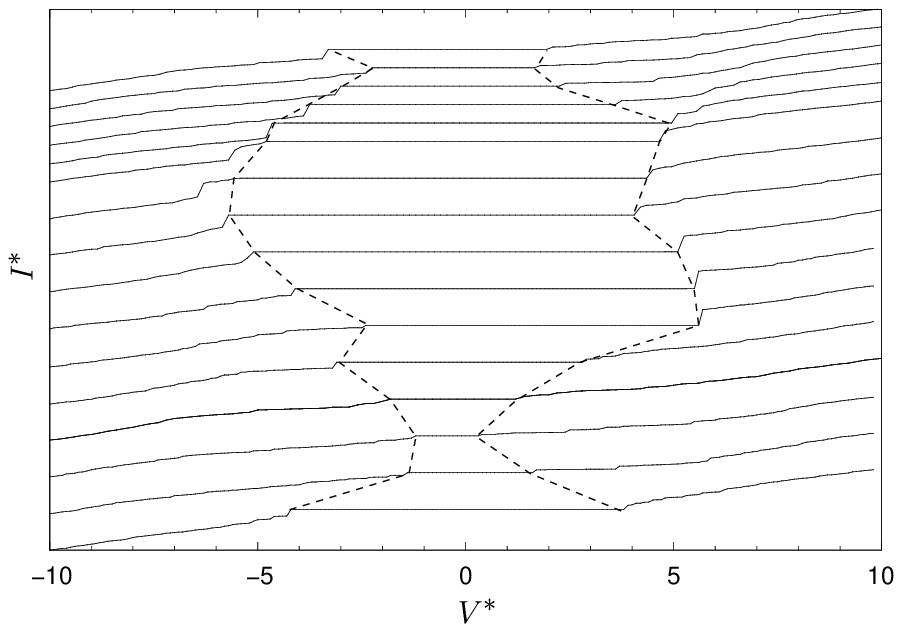}
\caption{\label{fig:arraygeom3a_gatecomp}Array of 4 nanoparticles, comparison
of IV curves for different gate voltages.
\textit{Upper} Linear fit of the ohmic part of the curves (dashed grey line) has the slope $G$ which is called \phem{overall conductance}  (denoted by the slope of the small triangle). The $V^{\ast}$-intercept of the fitted line is the \phem{offset voltage} $V_{\off}$ (denoted by the double-headed arrow). \textit{Lower} The current is multiplied by a factor and the curves are artificially offset from each other proportional to the applied gate voltage. The flat region in the middle always corresponds to Coulomb blockade, i.e. $I^{\ast} = 0$. The dashed line borders the Coulomb blockade region as a guide for the eye.} 
\centering
\end{figure}
In Fig.~\ref{fig:arraygeom3a_gatecomp} the IV characteristics of an array of 4 nanoparticles exemplify the typical IV characteristics of the array geometry. The various curves correspond to different gate voltages. We observe that for all array lengths two regions with different behaviours can be distinguished. For smaller bias voltages the curves exhibit a striking asymmetry with respect to reversal of the bias voltage. For example the threshold voltage is different for positive or negative bias voltages ($V^{\pm}_{T}$). This asymmetry is more pronounced for longer arrays and for curves with higher threshold voltages, especially if the $V^{\pm}_{T}$ exceed the offset voltage $V_{\off}$. The position, width and height of the steps in this bias voltage region and $V^{\pm}_{T}$ depend strongly on the gate voltage. For larger bias voltages, however, the IV curves shown in Fig.~\ref{fig:arraygeom3a_gatecomp} approximately coincide and become symmetric (with respect to reversal of the bias voltage).  For small bias voltages the number of many-particle states or the number of paths through the charge state space that take part in transport is smaller. So it is important \phem{which} states or paths actually participate. This is in turn influenced by the gate voltage since it shifts the energetic position of the charge states and determines therefore which paths are available. For bigger bias voltages many states or paths participate so it should be less important whether a certain state takes part or not: what we observe is their ``average'' contribution. That is why the curves for different gate voltages coincide for high bias voltages. It is also the reason for the disappearance of the asymmetry in the same region. A detailed discussion of the asymmetry can be found in the appendix \ref{sec:asym}.

\subsubsection{\label{subsubsec:cond}Overall conductance and offset voltage}
\begin{table}
\centering
\begin{tabular}{|c|c|c|c|c|} \hline
&\multicolumn{2}{c}{no relax. }&\multicolumn{2}{|c|}{fast
relax.}\\ \hline Z & $ V_{\off} $ & G & $ V_{\off} $ & G\\
\hline 2  & 2.3 & 0.12& 2.6 & 0.11\\ \hline 3 & 2.7 & 0.14 & 4.4 & 0.14\\
\hline 4 & 3.0 & 0.16 & 4.8 & 0.15\\ \hline 5 & 3.1 & 0.17 & & \\ \hline
\end{tabular}
\caption{  \label{tab:array} Offset voltage $V_{\off}$ and overall conductance
$G$ with respect to the number of nanoparticles $Z$} \centering
\end{table}
As already mentioned above, the IV curves shown in Fig.~\ref{fig:arraygeom3a_gatecomp} approximately coincide and become symmetric  for larger bias voltages. In that region the current increases linearly, apart from very small steps.  We fit this linear segment with a straight line and define its slope as the \phem{overall conductance} $G$. The \phem{offset voltage} $V_{\off}$, i.e. the $V^{\ast}$-intercept of the fitted line, can be thought of as a kind of ``mean'' threshold voltage. The actual threshold voltage of a single curve obviously depends on the gate voltage, as already mentioned. $G$ and $V_{\off}$ are approximately the same for positive and negative bias voltages. Our results for $G$ and $V_{\off}$ with respect to the number of nanoparticles $Z$ are compiled in the two left columns of Table \ref{tab:array} for the case without relaxation ($w^{rel} = 0$). We find that $G$ and $V_{\off}$ both increase with an increasing array length. 

The increase of $V_{\off}$ with an increasing array length has also been observed elsewhere \cite{Melsen97, Nguyen01} so it seems to be quite generic for arrays. One factor which contributes to this tendency is the decrease of the capacitances with increasing distance between the conductors. Especially the capacitances that couple the leads with the nanoparticles $\tilde{C}_{\alpha i} $, see eq.~\eqref{eq:cap3}, are important. The cpacitances extracted from the geometry with the help of ``FastCap'' \cite{White91} do not decrease linearly, like e.g.~for a simple parallel plate capacitor, but approximately exponential. This is reasonable since the nanoparticles partially screen the electric field. For two neighbouring nanoparticles that are in the middle of the array the difference between the coupling capacitances is small. So we have to apply a high bias voltage to create a potential difference between these two particles which permits a tunneling transition, see eq.~\eqref{eq:warr}. Therefore the threshold voltages and $V_{\off}$ increase with increasing array length. Concerning $G$ the change of the number of transport paths with increasing bias voltage plays an important role. This change is in turn determined by the change of the number of possible transitions. A certain transition between two neighbouring dots becomes possible when the potential gradient between the dots becomes high enough. If we assume that this gradient is roughly the bias voltage divided by the number of tunnel junctions then we should expect that $G$ decreases with increasing array length. On the other hand we ``grow'' the array by adding bigger nanoparticles with a higher density of one-electron states. Given that the number of extra electrons in the array is small a higher number of one-electron levels within the bias voltage window results in a higher current. This effect might overcompensate the reduced potential gradient which would result in the observed behaviour of $G$. To check this assertion we have artificially raised the level spacing on the last dot in the 3 particle array so that it is equal to the level spacing on the first dot: in this case we find that the overall conductance $G$ is only $0.1$ compared to $0.14$ with the original level spacing. So $G$ can indeed be lowered by raising the level spacing on the last dot. This supports our assertion. Obviously the offset voltage and the overall conductance can be tailored by the choice of the nanoparticle sizes. 

For the case of high relaxation rates $G$ and $V_{\off}$ are compiled in the two right columns of Table \ref{tab:array}. Comparing with the  case without relaxation we observe that the offset voltage is generally bigger if there is relaxation while the overall conductance is approximately the same. As argued in section \ref{subsubsec:relax} the current generally decreases with an increasing relaxation rate. This homogeneous downward shift of the IV curve correspondingly increases the offset voltage while the overall conductance remains unaltered.

\subsubsection{\label{subsubsec:giant}Giant step, giant Fano factor}
\begin{figure}
\centering
\includegraphics[width=0.49\textwidth]{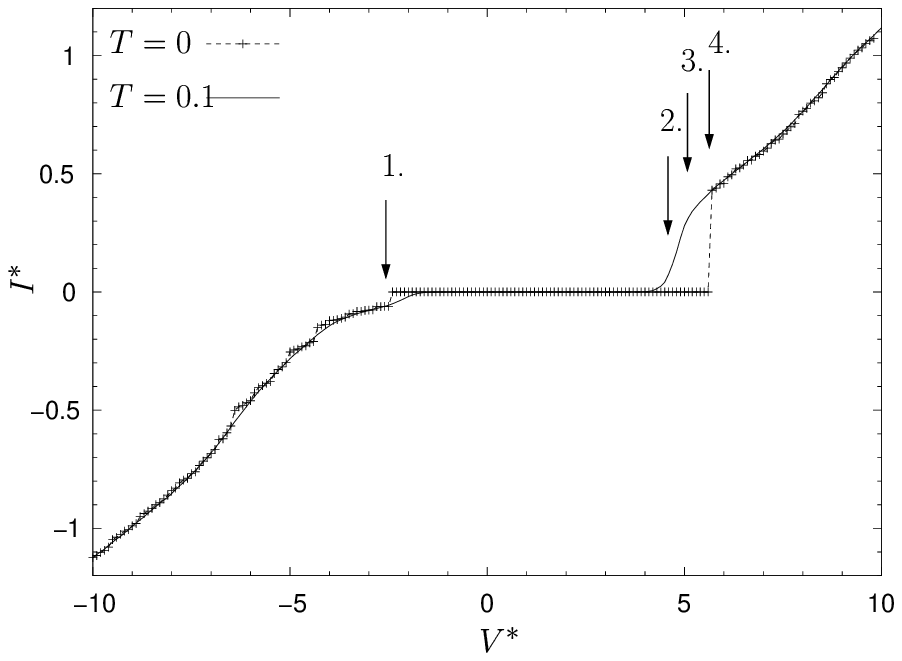}\\[-0.2cm]
{\includegraphics[width=0.49\textwidth]{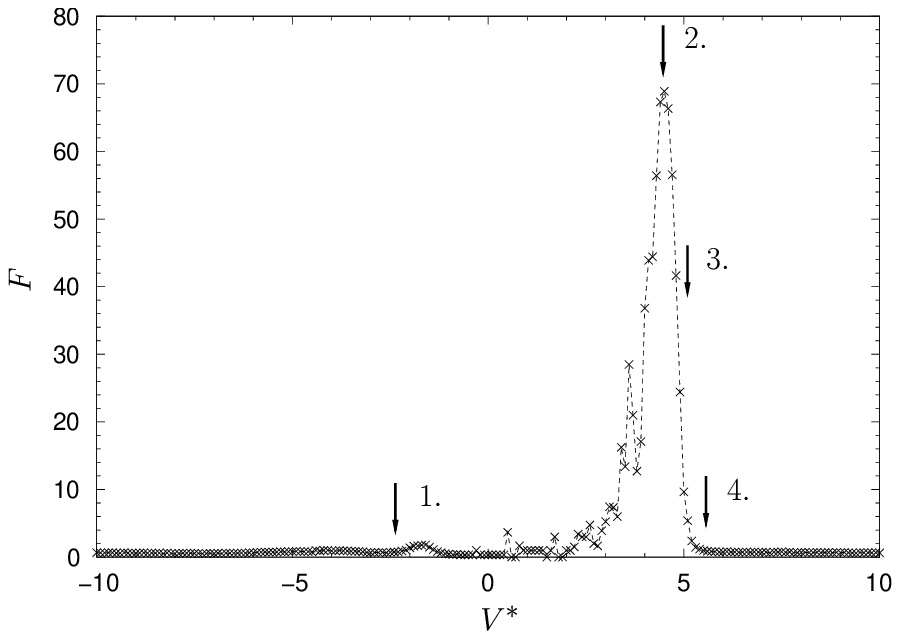}}
\caption{\label{fig:arraygeom3a_310_T01_IV.outs} 4 nanoparticle array, striking asymmetry in the IV characteristics (\textit{upper}) which is accompanied by a giant Fano factor (\textit{lower}). The temperature of $T=0.1$, for which also the Fano factor in the lower figure is calculated, equals about $280 K$.} 
\centering
\end{figure}
\begin{figure}
\centering
\includegraphics[width=0.47\textwidth]{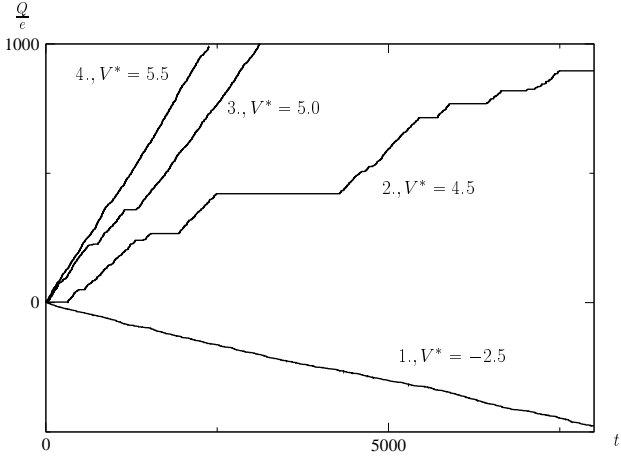}
\caption{ \label{fig:avalanche}Transferred charge vs.~time showing tunneling
avalanches, bias voltages correspond to those marked in
Fig.~\ref{fig:arraygeom3a_310_T01_IV.outs}:1. $V^{\ast} = -2.5$, 2.
$\rightarrow V^{\ast} = 4.5$,  3. $\rightarrow V^{\ast} = 5.0$,  4.
$\rightarrow V^{\ast} = 5.5$} 
\centering
\end{figure}
\begin{figure}
\centering
\includegraphics[width=0.3\textwidth]{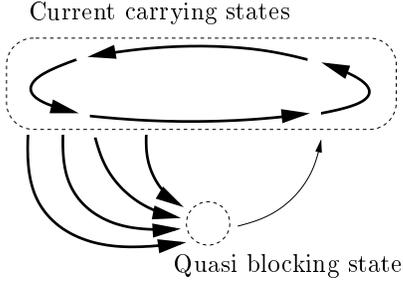}
\caption{\label{fig:blocking} Quasi-blocking state, the width of the arrows corresponds to the size of the transition rates. A quasi blocking state is characterized by numerous, big ingoing rates and few, small outgoing ones. In our case, the rates among the states that are involved in the charge transport are also much bigger than the rate leading out of the quasi blocking state.}
\centering
\end{figure}
As already stated above we observe that in the IV characteristics for several gate voltages the first current step to the left or to the right of the Coulomb blockade region is strikingly high. One example is shown in Fig.~\ref{fig:arraygeom3a_310_T01_IV.outs}. Furthermore the higher step is accompanied by a peaked giant Fano factor. To determine the origin of the big step in the current and the giant peak in the Fano factor we record the dynamics of the simulation, i.e. in Fig.~\ref{fig:avalanche} the transferred charge is plotted against the MC time for a single simulation run. This approach has already been used in a MC study of the electron transport properties in a different model\cite{Koch04} in which also a giant Fano factor is found. In Fig.~\ref{fig:avalanche} we have recorded the dynamics for the voltages marked in Fig.~\ref{fig:arraygeom3a_310_T01_IV.outs}. For the top of the peak marked with $2.$ we observe comparatively long periods without charge transfer and intermediate \phem{tunneling avalanches}. Looking at the definition of the Fano factor $F$, eq.~\eqref{eq:Fano}, there are two ways to conclude that the observed dynamics result in a super poissonian noise (i.e. $F>1$). On the one hand, one can regard the avalanches on a much bigger time scale as single, statistically independent transfer processes in which an effective charge bigger than $e$ is transported. Then one uses the Poisson value for the shot noise $2q \langle I \rangle$ with an effective charge $q$ bigger than $e$ which results in a Fano factor $F >1$. On the other hand, one can regard the avalanches as a \phem{bunching} of the tunneling processes and this positive correlation leads per definition to $F >1$. So both interpretations deduce the super poissonian shot noise from the observed dynamics. For bigger bias voltages, $3.$ and $4.$, the length of the periods without charge transfer is reduced so the Fano factor is reduced correspondingly.  The big step height can be understood analogously: during the periods without charge transfer a \phem{quasi-blocking state} is assumed , i.e. the sum of rates that lead out of this state is much smaller than the sum of the rates that lead into the same state, see Fig.~\ref{fig:blocking}. So if the state is visited the system rests there for a comparatively long time. If no rates led out of that state it would be a real \phem{blocking state} and the current would be zero since the dynamics of the system would ultimately end up in this state. The probability $P_{s}$ of a blocking state is therefore 1. In the shown case we find that a neutral many-electron state $(0,0,0,0)$ serves as a quasi-blocking state (respectively blocking state in the region of Coulomb blockade). If the system leaves it, current can flow. This current is however carried by \phem{other} states with much higher rates among them. Therefore the high step in the IV characteristics.  

\subsubsection{\label{subsubsec:NDC}Designable NDC effect}
\begin{figure}
\centering
\includegraphics[width=0.5\textwidth]{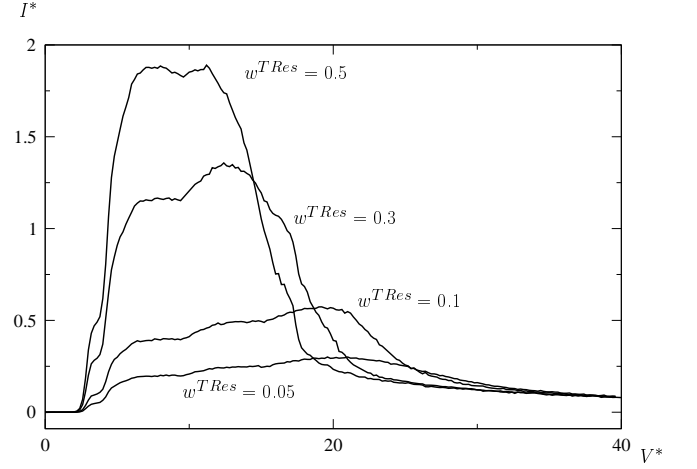}\\
\includegraphics[width=0.5\textwidth]{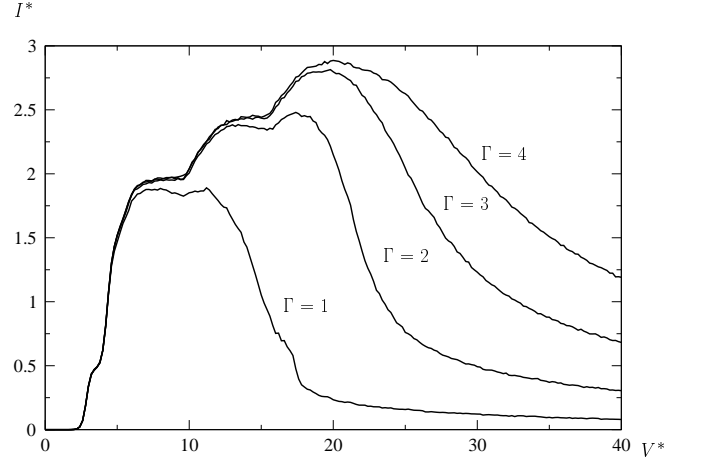}
\caption{\label{fig:NDCcomp} 2 nanoparticle array, NDC effect is stronger for higher tunneling rates between array and reservoirs ($w^{TRes}$) (\emph{upper}) whereas the NDC effect is attenuated by a broad $P(E)$ function i.e. for a large $\Gamma$ (\emph{lower})}
\centering
\end{figure}
\begin{figure}[h]
\centering
\includegraphics[width=0.25\textwidth]{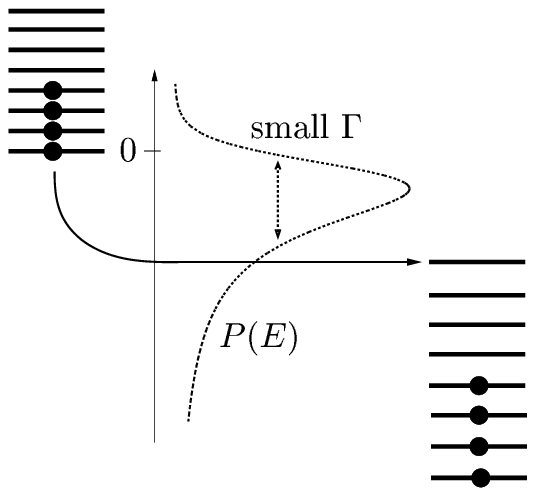}\\
\includegraphics[width=0.25\textwidth]{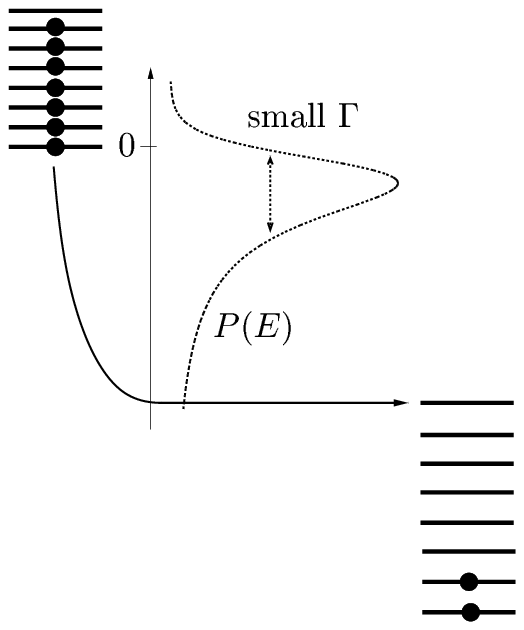}\\
\includegraphics[width=0.25\textwidth]{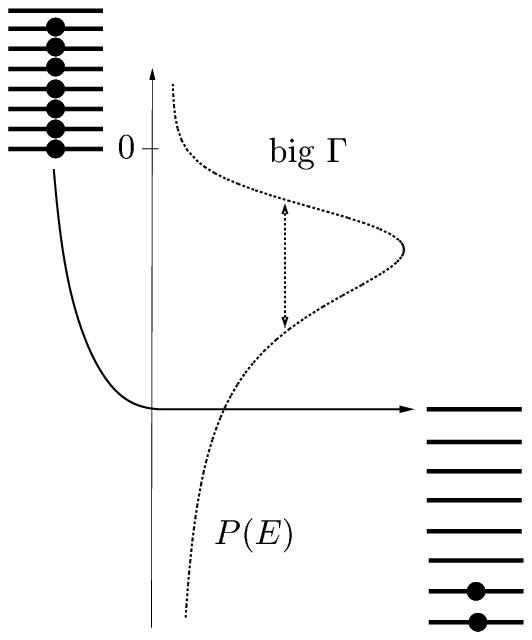}
\centering
\caption{\label{fig:NDCarr} 2 nanoparticle array, the energies $\left\lbrace l \cdot \Delta \varepsilon_{i} + e\Phi_{Ci} \right\rbrace_{l,i}$ are sketched. The $P(E)$ function determines the transition rates between the sketched levels.  \emph{Upper}: low bias voltage and small $\Gamma$. \emph{Middle}: high bias voltage and small $\Gamma$, the high bias voltage results in a large charge gradient between the nanoparticles which causes the large offset between the spectra. \emph{Lower}: high bias voltage and big $\Gamma$, a broad $P(E)$ function attenuates the NDC effect.} 
\end{figure}

So far we have modelled infinite spectra, see section \ref{subsec:monte}. Now we intentionally consider only a few levels on each nanoparticle. This is interesting since the electronic structure of small nanoparticles differs in general strongly from bulk material\cite{Barnett99}. We assume on each nanoparticle a finite ``band'' of discrete, equally spaced energy levels. The level spacing is estimated as given above, see eq.~ \eqref{eq:levelspacing}. Under these assumptions we find for an array of two nanoparticles a NDC which depends strongly on the ratio between transition rates within the array ($w^{TA}$) and the rates between array and reservoirs ($w^{TRes}$), see Fig.~\ref{fig:NDCcomp}. This ratio can be tuned by varying the distance between array and contacting reservoirs: the smaller the distance the higher the rate of tunneling between array and reservoirs ($w^{TRes}$). So this is an example of an effect which strongly depends on the geometry and which is therefore designable. The origin of the NDC can be understood by looking at the transition energies, see section \ref{subsec:transenerg}. In Fig.~\ref{fig:NDCarr} we sketch the energies $\left\lbrace l \cdot \Delta \varepsilon_{i} + e\Phi_{Ci} \right\rbrace_{l,i}$ for the two nanoparticles. For $T=0$ a transition between the nanoparticles can only happen if the difference between those energies is bigger than $\left ( C^{-1}_{ii}  - 2 C^{-1}_{i i \pm 1}  + C^{-1}_{i \pm 1 i \pm 1} \right )$, see eq.~\eqref{eq:deltaEarr}. For finite temperatures this restriction is softened by the coupling to the bosonic bath which results in the $P(E)$ function.  The offset between the sketched spectra is determined by the difference between the potentials $e\Phi_{Ci}$ which depend in turn on the charges on the nanoparticles. If the tunneling rates between array and reservoirs are sufficiently large a high charge gradient accumulates and therefore a large offset between the spectra occurs. Since the spectra are finite, the transition energies of possible transitions are big if the offset between the spectra is large. The corresponding value of the $P(E)$ function is small. Consequently the transition rates among the nanoparticles become smaller with increased voltage which results in the NDC. If the $P(E)$ function is broad, this effect is softened, see eq.~\eqref{eq:warr}. The higher the factor $\Gamma$ the broader the $P(E)$ function. Therefore the NDC effect is less distinct for higher values of $\Gamma$.

\section{\label{sec:concl}Conclusion}
We have shown that an improved MC algorithm can cope with the complexity of the electronic transport through nanoparticle arrays with discrete electronic spectra. Though we have considered linear arrays only we found a huge variety of transport functionalities. Of course, other geometries would also be worth studying, e.g. ring-shaped devices\cite{Shin98} can be used as charge storage elements. Since arrays can be made self-assembling, are robust against fabricational imperfections and show, as we found out, transport features that can be designed via the geometry, we consider them to be ideal building blocks for electronic architecture on the nanoscale.  

\begin{acknowledgements}
The authors gratefully acknowledge the financial support of the VW Foundation and the German National Merit Foundation. The authors would like to thank M. Holtschneider,  S. Jakobs, K. C. Nowack, M. Noyong, F. Reininghaus and M. R. Wegewijs for useful discussions.
\end{acknowledgements}

\appendix*
\section{\label{sec:asym}Asymmetry concerning reversal of bias voltage}
The occurence of IV curves which are asymmetric concerning the reversal of the bias voltage would be surprising if the transport was calculated within the Landauer-B\"uttiker formalism \cite{Landauer88,Buettiker86}. Within this approach the IV curves must be symmetric due to time reversal symmetry: for each wave which travels through the system from left to right there is a time reversed one with the opposite direction of propagation. But the Landauer-B\"uttiker formalism can only be applied for coherent transport without interaction. Both suppositions are violated in our case: we assume that the phase information is lost at every tunneling event and we include Coulomb interaction, see sections \ref{subsec:hamil} and \ref{subsec:transport}. Therefore, in our case, asymmetry is present in general unless special symmetries impose symmetric curves. We will discuss now two symmetries that are actually relevant in our case. By explicitly showing how these symmetries are broken by the Coulomb interaction and the distribution of the density of states (inverse level spacing) the occurence of asymmetric IV curves is rendered plausible. We adopt the framework of the orthodox theory for these considerations since the basic mechanisms can be understood by looking at charge states alone.

\subsection{\label{subsec:charge sym}Particle-hole symmetry}
This symmetry means that two paths through the charge state space with opposite charges are equivalent. Equivalent here means that they appear at the same absolute value but opposite sign of the bias voltage and that the rates for corresponding transitions are equal. E.g.~for a 2 nanoparticle array this means that if the path $(-1,-2) \rightarrow (-2,-2) \rightarrow (-1,-3) \rightarrow (-1,-2)$ appears at $V=x$ then the path $(1,2) \rightarrow (2,2) \rightarrow (1,3) \rightarrow (1,2)$ appears at $V=-x$ and the rates for corresponding transitions are equal. This symmetry is present if there is no coupling to the gate and if there are no background charges. To conclude this we have to look at eq.~\eqref{eq:bgcharge} for the potentials $\un{\Phi}_{C}$ on the nanoparticles: $ \un{\Phi}_{C} = C^{-1}_{c} \left (\un{Q}_{C} + \un{Q}'_{C}  + \un{\tilde{Q}}^{bg}_{C}  \right ) $. If there is no coupling to the gate the polarization charges on the nanoparticles are reversed if the bias voltage is reversed: $V \rightarrow -V \Rightarrow  \un{Q}'_{C} \rightarrow -\un{Q}'_{C}$. If we reverse the charges $\un{Q}_{C}$ at the same time and there are no background charges $\un{\tilde{Q}}^{bg}_{C}$, the potentials  $\un{\Phi}_{C}$  are reversed: $V \rightarrow -V  , \un{Q}_{C} \rightarrow -\un{Q}_{C} \Rightarrow  \un{\Phi}_{C} \rightarrow -\un{\Phi}_{C}$. Since the potentials  $\un{\Phi}_{C}$ determine the transition rates two charge states with opposite charges are equivalent in the sense explained above. Obviously finite background charges or coupling to the gate break this symmetry. 

\subsection{\label{subsec:mirrorsym}Inversion symmetry}
This symmetry means that two paths through the charge state space which are mirror images of each other are equivalent where the mirror plane shall be situated in the middle of the array and equivalent is meant in the sense explained above. E.g. for a 2 nanoparticle array this means that if the path $(-1,-2) \rightarrow (-2,-2) \rightarrow (-1,-3) \rightarrow (-1,-2)$ appears at $V=x$ then the path $(-2,-1) \rightarrow (-2,-2) \rightarrow (-3,-1) \rightarrow (-2,-1)$ appears at $V=-x$ and the rates for corresponding transitions are equal. This symmetry is obviously present if the whole setup is symmetric with respect to a mirror plane in the middle of the array. Both an asymmetric capacitance matrix and an asymmetric distribution of the density of states on the nanoparticles break this symmetry.

\subsubsection{\label{subsubsec:mirrorsym_capmat}Asymmetric capacitance matrix}
We look at an array of 2 nanoparticles with different sizes, the left nanoparticle ($A$) shall be smaller than the other ($B$). The capacitative coupling (between nodes) shall be negligible so that the capacitance matrix is diagonal. Then each nanoparticle can be characterized by a single capacitance, $C_{A}$ and $C_{B}$ respectively. Since nanoparticle $A$ is smaller, it holds $C_{A} \ll C_{B}$. The opposite holds for the charging energies $E_{C_{A}} \gg E_{C_{B}}$ so more energy is needed to charge nanoparticle $A$. Now it is clear that the paths $(0,0) \rightarrow (-1,0) \rightarrow (0,-1) \rightarrow (0,0)$ and $(0,0) \rightarrow (0,-1) \rightarrow (-1,0) \rightarrow (0,0)$ are not equivalent. The asymmetric capacitance matrix breaks the inversion symmetry.

\subsubsection{\label{subsubsec:mirrorsym_DOS}Asymmetric distribution of the density of states}
The capacitance matrix of the 2 nanoparticle array shall now be symmetric and still diagonal, so the charging energy is the same for both nanoparticles. The density of states, however, shall be $D_{A}$ on the left and $D_{B}$ on the right nanoparticle.  (This corresponds to different level spacings in our model.)  The density of states in both reservoirs shall be $D_{Res}$. The energy change for the three tunneling transitions that appear in the path $(0,0) \rightarrow (-1,0) \rightarrow (0,-1) \rightarrow (0,0)$ shall be denoted by $\Delta E_{1}$,$\Delta E_{2}$ and $\Delta E_{3}$. (The index denotes the number of the transition e.g.~$\Delta E_{1}$ is the energy difference occuring at the transition $(0,0) \rightarrow (-1,0)$.) The corresponding  transition rates are $w_{1}$, $w_{2}$ and $w_{3}$. For zero temperature the orthodox theory rates are:
\begin{eqnarray} 
w_{1} &=& \frac{2 \pi}{\hbar} \left |t^{Res} \right |^2 (-\Delta E_{1}) \Theta (-\Delta E_{1}) D_{Res} D_{A} \\ 
w_{2} &=& \frac{2 \pi}{\hbar} \left |t^{A} \right |^2 (-\Delta E_{2}) \Theta (-\Delta E_{2}) D_{A}D_{B} \\ 
w_{3} &=& \frac{2 \pi}{\hbar} \left |t^{Res} \right |^2(-\Delta E_{3}) \Theta (-\Delta E_{3}) D_{B}D_{Res} 
\end{eqnarray}
If we reverse the bias voltage and look at the path  $(0,0) \rightarrow (0,-1) \rightarrow (-1,0) \rightarrow (0,0)$,  we get different energy differences $\Delta E'_{1}$,$\Delta E'_{2}$ and $\Delta E'_{3}$ and rates  $w'_{1}$, $w'_{2}$ and $w'_{3}$. (Note that here e.g.~$\Delta E'_{1}$ is the energy difference occuring at the transition $(0,0) \rightarrow (0,-1)$.) Since the capacitance matrix is symmetric, it holds $\Delta E'_{i} =\Delta E_{i} \: \forall i$. So we get: 
\begin{eqnarray} 
w'_{1} &=& \frac{2 \pi}{\hbar} \left |t^{Res} \right |^2 (-\Delta E_{1}) \Theta (-\Delta E_{1}) D_{Res} D_{B} \\ 
w'_{2} &=& \frac{2 \pi}{\hbar} \left |t^{A} \right |^2 (-\Delta E_{2}) \Theta (-\Delta E_{2}) D_{A}D_{B} \\ 
w'_{3} &=& \frac{2 \pi}{\hbar} \left |t^{Res} \right |^2(-\Delta E_{3}) \Theta (-\Delta E_{3}) D_{A}D_{Res} 
\end{eqnarray}
Obviously the rates $\left\lbrace w_{i} \right\rbrace_{i}$ are different from the rates $\left\lbrace w'_{i} \right\rbrace_{i}$ due to the different density of states $\left\lbrace D_{i} \right\rbrace_{i}$ so the considered paths are not equivalent. An asymmetric distribution of the densities of states breaks the inversion symmetry.


\begin{thebibliography}{10}

\bibitem{Niemeyer04}
C.M.\ Niemeyer and C.A.\ Mirkin, editors.
\newblock {\em Nanobiotechnology}, chapter~17.
\newblock VCH, 2004.

\bibitem{Zhang94}
Y.\ Zhang and N.C.\ Seeman.
\newblock {\em J. Am. Chem. Soc.}, 116:1661, 1994.

\bibitem{Noyong03}
C.M.\ Niemeyer, B.\ Ceyhan, M.\ Noyong, and U.\ Simon.
\newblock {\em Biochem. and Biophys. Res. Comm.}, 311:995, 2003.

\bibitem{Berven02}
C.A.\ Berven, M.N.\ Wybourne, J.\ Clarke, L.\ Longstreth, and J.E.\ Hutchison.
\newblock {\em J. Appl. Phys.}, 92:4513, 2002.

\bibitem{Noyong03b}
M.\ Noyong, K.\ Gloddeck, and U.\ Simon.
\newblock Assembly of gold nanoparticles on dna strands.
\newblock In {\em Mat. Res. Soc. Symp. Proc.}, volume 735, pages 153--158,
  2003.

\bibitem{Wasshuber01}
C.\ Wasshuber.
\newblock {\em Computational Single-Electronics}.
\newblock Springer, 2001.  
 
\bibitem{Bylander05}
J.\ Averin, T.\ Duty, and P.\ Delsing
\newblock {\em Nature}, 434:285, 2005. 
   
\bibitem{Averin05}
V.\ Averin
\newblock {\em Nature}, 434:285, 2005.

\bibitem{White91}
K.~\ Nabors and J.~\ White.
\newblock Fastcap: A multipole-accelerated 3-d capacitance extraction program.
\newblock {\em IEEE Transactions on Computer-Aided Design}, 10:1447, 1991.

\bibitem{Amman89}
M.\ Amman, E.\ Ben-Jacob, and K.\ Mullen.
\newblock {\em Phys.\ Lett.\ A}, 135:390, 1989.

\bibitem{Amman89b}
M.\ Amman, E.\ Ben-Jacob, and K.\ Mullen.
\newblock {\em Phys.\ Lett.\ A}, 142:431, 1989.

\bibitem{Geigenmueller89}
U.\ Geigenm\"uller and G.~Sch\"on.
\newblock {\em Europhys.\ Lett.}, 10:765, 1989.

\bibitem{Bakhvalov89}
N.S.\ Bakhvalov, G.S.\ Kazacha, K.H.\ Likharev, and S.I. Serdyukova.
\newblock {\em Sov. Phys. JETP}, 68:581, 1989.

\bibitem{Middleton93}
A.~A.\ Middleton and N.S.\ Wingreen.
\newblock {\em Phys.\ Rev.\ Lett.}, 71:3198, 1993.

\bibitem{Melsen97}
J.~A.\ Melsen, U.\ Hanke, H.-O.\ M\"uller, and K.-A.\ Chao.
\newblock {\em Phys.\ Rev.\ B}, 55:10638, 1997.

\bibitem{Nguyen01}
V.L.\ Nguyen, T.D.\ Nguyen, and H.N.\ Nguyen.
\newblock {\em Phys.\ Lett.\ A}, 291:150, 2001.

\bibitem{Likharev99}
K.\ Likharev.
\newblock Single-electron devices and their applications.
\newblock In {\em Proc. IEEE}, volume~87, pages 606--632, 1999.

\bibitem{Mahan90}
G.\ Mahan.
\newblock {\em Many-Particle Physics}.
\newblock Plenum Press, New York, 1990.

\bibitem{Nazarov92}
G.~L.\ Ingold and Yu.~V.\ Nazarov.
\newblock Charge tunneling rates in ultrasmall junctions.
\newblock In H.\ Grabert and M.H.\ Devoret, editors, {\em Single Charge
  Tunneling}, chapter~2. Plenum Press, New York, 1992.

\bibitem{Devoret94}
H.~\ Devoret, D.~\ Esteve, and C.~\ Urbina.
\newblock Single electron phenomena in metallic nanostructures.
\newblock In E.\ Akkermans, G.\ Montambaux, J.L.\ Pichard, and J.\ Zinn-Justin,
  editors, {\em Les Houches Summer School, Session LXI}, Amsterdam, 1994.
  North-Holland.

\bibitem{Whan96}
C.~B.~\ Whan, J.~\ White, and T.P.\ Orlando.
\newblock {\em Appl.\ Phys.\ Lett.}, 68:2996, 1996.

\bibitem{Halperin86}
W.P.\ Halperin.
\newblock {\em Rev.\ Mod.\ Phys.}, 58:533, 1986.

\bibitem{Barnett99}
R.~N.\ Barnett, C.L.\ Cleveland, H.\ H\"akinnen, W.D.\ Luedtke, C.\ Yannouless,
  and U.\ Landmann.
\newblock {\em Eur.\ Phys.\ J.\ D}, 9:95, 1999.

\bibitem{Zhang04}
H.\ Zhang, U.\ Hartmann, and G.\ Schmid.
\newblock {\em Appl.\ Phys.\ Lett.}, 84:1543, 2004.

\bibitem{Schoeller97}
H.\ Schoeller.
\newblock Transport theory of interacting quantum dots.
\newblock In L.L. Sohn, L.P. Kouwenhoven, and G.~Sch\"on, editors, {\em
  Mesoscopic Electron Transport}. Kluwer, Dordrecht, 1997.

\bibitem{Honerkamp94}
J.\ Honerkamp.
\newblock {\em Stochastic Dynamical Systems}.
\newblock VCH, 1994.

\bibitem{Binder97}
K.\ Binder.
\newblock {\em Rep. Prog. Phys.}, 60:487, 1997.

\bibitem{Korotkov94}
A.~N.\ Korotkov.
\newblock {\em Phys.\ Rev.\ B}, 49:10381, 1994.

\bibitem{Averin91}
D.~V.\ Averin, A.N.\ Korotkov, and K.K.\ Likharev.
\newblock {\em Phys.\ Rev.\ B}, 44:6199, 1991.

\bibitem{Averin90}
D.~V.\ Averin and A.N.\ Korotkov.

\bibitem{Koch04}
J.~Koch and F.~von Oppen.
\newblock cond-mat/0409667, 2004.

\bibitem{Shin98}
M.\ Shin, S.\ Lee, K.W.\ Park, and E.\ Lee.
\newblock {\em Phys.\ Rev.\ Lett.}, 80:5774, 1998.

\bibitem{Landauer88}
R.\ Landauer.
\newblock {\em IBM J. Res. Dev.}, 32:306, 1988.

\bibitem{Buettiker86}
M.\ B\"uttiker.
\newblock {\em Phys.\ Rev.\ Lett.}, 57:1761, 1986.




\end{thebibliography}
\end{document}